________________________________________________________________________

\documentstyle[12pt]{article}
\title{{\normalsize{{\hskip 8.5cm} BIHEP-TH-94-41}}\\[-7mm]
{\normalsize{{\hskip 8.5cm} December,~~~1994}}\\
$R$ Matrix for $U_{q}E_{7}$}

\author{Jin Bai-Qi and Ma Zhong-Qi\\
{\small Institute of High Energy Physics, P.O.Box 918(4), Beijing
100039, P. R. of China}}

\date{}
\begin{document}
\maketitle

\vspace{20mm}

\begin{abstract}
The quantum Clebsch-Gordan coefficients and the explicit form of the
$\breve{R}_{q}$ matrix related with the minimal representation of
the quantum enveloping algebra $U_{q}E_{7}$ are calculated in this paper.
\end{abstract}

\newpage
\noindent
{\bf 1. INTRODUCTION}

\vspace{2mm}
Recently, quantum groups $^{1}$ have drawn an increasing attention by
both physicists and mathematicians. A quantum group is introduced as the
non-commutative and non-cocommutative Hopf algebra ${\cal A}$ obtained
by continuous deformations of the Hopf algebra of the function of
a Lie group. The associative algebra ${\cal A}$ is freely generated by
non-commutating matrix entries $T^{a}_{~b}$, satisfying the so-called
$RTT=TTR$ relation $^{2}$:
$$\left(\hat{R}_{q}\right)^{ab}_{~rs}~T^{r}_{~c}~T^{s}_{~d}~=~
T^{a}_{~r}~T^{b}_{~s}~\left(\hat{R}_{q}\right)^{rs}_{~cd}
\eqno (1.1) $$

\noindent
where $\hat{R}_{q}$ matrix is a solution of the simple Yang-Baxter
equation $^{3}$ related with the minimal
representation of the corresponding quantum enveloping algebra.

The $\hat{R}_{q}$ matrices are well known for the quantum classical
Lie enveloping algebras $^{4}$. Based on those solutions and (1.1)
the quantum groups $A_{\ell}(q)$, $B_{\ell}(q)$, $C_{\ell}(q)$
and $D_{\ell}(q)$, and their bicovariant differential calculus
were studied $^{5,6}$. In order to generalize the concept of the
quantum group into the quantum exceptional groups, one has to
calculate the $\hat{R}_{q}$ matrices for the quantum exceptional
enveloping algebras firstly. Kuniba $^{7}$ calculated the quantum
Clebsch-Gordan ($q$-CG) coefficients for the minimal representation of
$U_{q}G_{2}$. Although he did not list the explicit elements of
the $\hat{R}_{q}$ matrix in his paper, it is not hard to calculate
them from the $q$-CG coefficients. The explicit $\hat{R}_{q}$ matrices
for $U_{q}F_{4}$ and $U_{q}E_{6}$ were listed in Refs. [8] and [9],
but partly $^{9}$ for $U_{q}E_{7}$. The most complicated $56 \times
56$ submatrix did not given there. In the direct product space of
two minimal representations spaces there are 56 states with weight zero.
It is a very complicated task to combine them into orthogonal
bases belonging to four irreducible representations, respectively.
In the present paper we use a systematic method $^{10}$ to calculate
both the representation matrix and the $q$-CG matrix simultaneously
in terms of the Mathematica program, and succeed in obtaining the
$56 \times 56$ submatrix of the $\hat{R}_{q}$ matrix.

The plan of this paper is as follows. In Sec. 2, we introduce our notation.
The main results of $q$-CG coefficients and the $\hat{R}_{q}$ matrix
are listed in Sec.3 and Sec.4.

\vspace{10mm}
\noindent
{\bf 2. NOTATIONS}

\begin{center}
\setlength{\unitlength}{1mm}
\begin{picture}(50,22)(0,-15)
\put(0,0){\circle{2}}
\put(0,-4){\makebox(0,0){1}}
\put(1,0){\line(1,0){8}}
\put(10,0){\circle{2}}
\put(10,-4){\makebox(0,0){2}}
\put(11,0){\line(1,0){8}}
\put(20,0){\circle{2}}
\put(20,-4){\makebox(0,0){3}}
\put(21,0){\line(1,0){8}}
\put(30,0){\circle{2}}
\put(30,-4){\makebox(0,0){4}}
\put(31,0){\line(1,0){8}}
\put(40,0){\circle{2}}
\put(40,-4){\makebox(0,0){5}}
\put(41,0){\line(1,0){8}}
\put(50,0){\circle{2}}
\put(50,-4){\makebox(0,0){6}}
\put(20,8){\circle{2}}
\put(24,8){\makebox(0,0){7}}
\put(20,1){\line(0,1){6}}
\put(28,-10){\makebox(0,0){{\bf Fig.1.1}. Dynkin diagram for $E_{7}$.}}
\end{picture}
\end{center}

\begin{center}
\setlength{\unitlength}{1.0mm}
\begin{picture}(70,165)(0,-10)

\put(24,150){\framebox(18,6){$000~0010$}}
\put(24,140){\framebox(18,6){$000~01\bar{1}0$}}
\put(24,130){\framebox(18,6){$000~1\bar{1}00$}}
\put(24,120){\framebox(18,6){$001~\bar{1}000$}}
\put(24,110){\framebox(18,6){$01\bar{1}~0001$}}
\put(12,100){\framebox(18,6){$1\bar{1}0~0001$}}
\put(36,100){\framebox(18,6){$010~000\bar{1}$}}
\put(12,90){\framebox(18,6){$\bar{1}00~0001$}}
\put(36,90){\framebox(18,6){$1\bar{1}1~000\bar{1}$}}
\put(12,80){\framebox(18,6){$\bar{1}01~000\bar{1}$}}
\put(36,80){\framebox(18,6){$10\bar{1}~1000$}}
\put(12,70){\framebox(18,6){$\bar{1}1\bar{1}~1000$}}
\put(36,70){\framebox(18,6){$100~\bar{1}100$}}
\put(0,60){\framebox(18,6){$0\bar{1}0~1000$}}
\put(24,60){\framebox(18,6){$\bar{1}10~\bar{1}100$}}
\put(48,60){\framebox(18,6){$100~0\bar{1}10$}}
\put(0,50){\framebox(18,6){$0\bar{1}1~\bar{1}100$}}
\put(24,50){\framebox(18,6){$\bar{1}10~0\bar{1}10$}}
\put(48,50){\framebox(18,6){$100~00\bar{1}0$}}
\put(0,40){\framebox(18,6){$00\bar{1}~0101$}}
\put(24,40){\framebox(18,6){$0\bar{1}1~0\bar{1}10$}}
\put(48,40){\framebox(18,6){$\bar{1}10~00\bar{1}0$}}
\put(0,30){\framebox(18,6){$000~010\bar{1}$}}
\put(24,30){\framebox(18,6){$00\bar{1}~1\bar{1}11$}}
\put(48,30){\framebox(18,6){$0\bar{1}1~00\bar{1}0$}}
\put(0,20){\framebox(18,6){$000~1\bar{1}1\bar{1}$}}
\put(24,20){\framebox(18,6){$000~\bar{1}011$}}
\put(48,20){\framebox(18,6){$00\bar{1}~10\bar{1}1$}}

\put(0,6){\framebox(18,6){$001~\bar{1}01\bar{1}$}}
\put(24,6){\framebox(18,6){$000~10\bar{1}\bar{1}$}}
\put(48,6){\framebox(18,6){$000~\bar{1}1\bar{1}1$}}

\put(0,16){\dashbox{1}(66,0){}}
\put(33,150){\line(0,-1){4}}
\put(34,148){\makebox(0,0)[l]{{\small 6}}}
\put(33,140){\line(0,-1){4}}
\put(34,138){\makebox(0,0)[l]{{\small 5}}}
\put(33,130){\line(0,-1){4}}
\put(34,128){\makebox(0,0)[l]{{\small 4}}}
\put(33,120){\line(0,-1){4}}
\put(34,118){\makebox(0,0)[l]{{\small 3}}}
\put(24,110){\line(0,-1){4}}
\put(23,108){\makebox(0,0)[r]{{\small 2}}}
\put(42,110){\line(0,-1){4}}
\put(43,108){\makebox(0,0)[l]{{\small 7}}}
\put(21,100){\line(0,-1){4}}
\put(22,98){\makebox(0,0)[l]{{\small 1}}}
\put(45,100){\line(0,-1){4}}
\put(46,98){\makebox(0,0)[l]{{\small 2}}}
\put(29,100){\line(2,-1){8}}
\put(33,99){\makebox(0,0)[b]{{\small 7}}}
\put(21,90){\line(0,-1){4}}
\put(22,88){\makebox(0,0)[l]{{\small 7}}}
\put(45,90){\line(0,-1){4}}
\put(46,88){\makebox(0,0)[l]{{\small 3}}}
\put(35,90){\line(-2,-1){8}}
\put(32,89){\makebox(0,0)[b]{{\small 1}}}
\put(21,80){\line(0,-1){4}}
\put(22,78){\makebox(0,0)[l]{{\small 3}}}
\put(45,80){\line(0,-1){4}}
\put(46,78){\makebox(0,0)[l]{{\small 4}}}
\put(35,80){\line(-2,-1){8}}
\put(32,79){\makebox(0,0)[b]{{\small 1}}}
\put(12,70){\line(0,-1){4}}
\put(13,68){\makebox(0,0)[l]{{\small 2}}}
\put(30,70){\line(0,-1){4}}
\put(31,68){\makebox(0,0)[l]{{\small 4}}}
\put(36,70){\line(0,-1){4}}
\put(37,68){\makebox(0,0)[l]{{\small 1}}}
\put(54,70){\line(0,-1){4}}
\put(55,68){\makebox(0,0)[l]{{\small 5}}}
\put(9,60){\line(0,-1){4}}
\put(10,58){\makebox(0,0)[l]{{\small 4}}}
\put(33,60){\line(0,-1){4}}
\put(34,58){\makebox(0,0)[l]{{\small 5}}}
\put(25,60){\line(-2,-1){8}}
\put(21,59){\makebox(0,0)[b]{{\small 2}}}
\put(49,60){\line(-2,-1){8}}
\put(45,59){\makebox(0,0)[b]{{\small 1}}}
\put(57,60){\line(0,-1){4}}
\put(58,58){\makebox(0,0)[l]{{\small 6}}}
\put(9,50){\line(0,-1){4}}
\put(10,48){\makebox(0,0)[l]{{\small 3}}}
\put(33,50){\line(0,-1){4}}
\put(34,48){\makebox(0,0)[l]{{\small 2}}}
\put(17,50){\line(2,-1){8}}
\put(21,49){\makebox(0,0)[b]{{\small 5}}}
\put(41,50){\line(2,-1){8}}
\put(45,49){\makebox(0,0)[b]{{\small 6}}}
\put(57,50){\line(0,-1){4}}
\put(58,48){\makebox(0,0)[l]{{\small 1}}}
\put(9,40){\line(0,-1){4}}
\put(10,38){\makebox(0,0)[l]{{\small 7}}}
\put(33,40){\line(0,-1){4}}
\put(34,38){\makebox(0,0)[l]{{\small 3}}}
\put(17,40){\line(2,-1){8}}
\put(21,39){\makebox(0,0)[b]{{\small 5}}}
\put(41,40){\line(2,-1){8}}
\put(45,39){\makebox(0,0)[b]{{\small 6}}}
\put(57,40){\line(0,-1){4}}
\put(58,38){\makebox(0,0)[l]{{\small 2}}}
\put(9,30){\line(0,-1){4}}
\put(10,28){\makebox(0,0)[l]{{\small 5}}}
\put(33,30){\line(0,-1){4}}
\put(34,28){\makebox(0,0)[l]{{\small 4}}}
\put(25,30){\line(-2,-1){8}}
\put(21,29){\makebox(0,0)[b]{{\small 7}}}
\put(41,30){\line(2,-1){8}}
\put(45,29){\makebox(0,0)[b]{{\small 6}}}
\put(57,30){\line(0,-1){4}}
\put(58,28){\makebox(0,0)[l]{{\small 3}}}
\put(6,20){\line(0,-1){8}}
\put(7,18){\makebox(0,0)[l]{{\small 4}}}
\put(15,20){\line(3,-2){12}}
\put(16,19){\makebox(0,0)[t]{{\small 6}}}
\put(27,20){\line(-3,-2){12}}
\put(26,19){\makebox(0,0)[t]{{\small 7}}}
\put(39,20){\line(3,-2){12}}
\put(40,19){\makebox(0,0)[t]{{\small 6}}}
\put(51,20){\line(-3,-2){12}}
\put(50,19){\makebox(0,0)[t]{{\small 7}}}
\put(60,20){\line(0,-1){8}}
\put(61,18){\makebox(0,0)[l]{{\small 4}}}

\put(43,153){\makebox(0,0)[l]{{\small 28}}}
\put(43,143){\makebox(0,0)[l]{{\small 27}}}
\put(43,133){\makebox(0,0)[l]{{\small 26}}}
\put(43,123){\makebox(0,0)[l]{{\small 25}}}
\put(43,113){\makebox(0,0)[l]{{\small 24}}}
\put(11,103){\makebox(0,0)[r]{{\small 23}}}
\put(55,103){\makebox(0,0)[l]{{\small 22}}}
\put(11,93){\makebox(0,0)[r]{{\small 21}}}
\put(55,93){\makebox(0,0)[l]{{\small 20}}}
\put(11,83){\makebox(0,0)[r]{{\small 19}}}
\put(55,83){\makebox(0,0)[l]{{\small 18}}}
\put(11,73){\makebox(0,0)[r]{{\small 17}}}
\put(55,73){\makebox(0,0)[l]{{\small 16}}}
\put(-1,64){\makebox(0,0)[r]{{\small 15}}}
\put(23,64){\makebox(0,0)[r]{{\small 14}}}
\put(67,64){\makebox(0,0)[l]{{\small 13}}}
\put(-1,54){\makebox(0,0)[r]{{\small 12}}}
\put(23,54){\makebox(0,0)[r]{{\small 11}}}
\put(67,54){\makebox(0,0)[l]{{\small 10}}}
\put(-1,44){\makebox(0,0)[r]{{\small 9}}}
\put(23,44){\makebox(0,0)[r]{{\small 8}}}
\put(67,44){\makebox(0,0)[l]{{\small 7}}}
\put(-1,34){\makebox(0,0)[r]{{\small 6}}}
\put(23,34){\makebox(0,0)[r]{{\small 5}}}
\put(67,34){\makebox(0,0)[l]{{\small 4}}}
\put(-1,24){\makebox(0,0)[r]{{\small 3}}}
\put(23,24){\makebox(0,0)[r]{{\small 2}}}
\put(67,24){\makebox(0,0)[l]{{\small 1}}}
\put(-1,9){\makebox(0,0)[r]{{\small $\bar{1}$}}}
\put(23,9){\makebox(0,0)[r]{{\small $\bar{2}$}}}
\put(67,9){\makebox(0,0)[l]{{\small $\bar{3}$}}}

\put(33,-2){\makebox(0,0){{\bæ Fig.2}® Up-half block weight
diagram foò thå highest}}
\put(33,-7){\makebox(0,0){ weight representatioî ${\bf \lambda}_{6}$ of
$U_{q}E_{7}$}}

\end{picture}
\end{center}

The Dynkin diagram for the exceptional group $E_{7}$ is showed in
Fig.1. There are seven simple roots ${\bf r}_{j}$ and seven
fundamental dominant weights ${\bf \lambda}_{j}$ in $E_{7}$ that
are related by the Cartan matrix $A_{ij}$:
$${\bf r}_{j}~=~\displaystyle \sum_{i=1}^{7}~{\bf \lambda}_{i}~A_{ij}
\eqno (2.1) $$

An irreducible representation is denoted by the highest weight
${\bf M}$ that is an non-negative integral combination of the
fundamental weights ${\bf \lambda}_{j}$, and the states in the
representation by weights ${\bf m}$ that are integral combinations
of ${\bf \lambda}_{j}$. The minimal representation of $E_{7}$
is ${\bf \lambda}_{6}$ that is 56 dimensional. Since the block
weight diagram of the minimal representation is symmetry up and
down, only up-half part of the diagram is given in Fig.2.

To make notation simpler we enumerate the states by one index $a$:
$$a~=~28,~ 27,~ \cdots~,~ 2,~ 1,~ \bar{1},~ \bar{2},~ \cdots,~ \bar{28},
{}~~~~~\bar{a}~\equiv~-~a \eqno (2.2) $$

\noindent
The index $a$ is written by the blocks in Fig.2. Denote by ${\bf m}_{a}$ the
weight of the state $a$, and by $h_{a}$ the height of the state. The
weights of the states are filled in the blocks of Fig.2. For example,
${\bf m}_{28}={\bf \lambda}_{6}$, and ${\bf m}_{23}={\bf \lambda}_{1}
-{\bf \lambda}_{2}+ {\bf \lambda}_{7}$. The heights of the states
are calculated by:
$${\bf M}~-~{\bf m}_{a}~=~\displaystyle \sum_{j=1}^{7}~c_{j}~{\bf r}_{j},~~~~
h_{a}~=~\displaystyle \sum_{j=1}^{7}~c_{j} $$

\noindent
and listed as follows:
$$\begin{array}{lll}
h_{28}=0,&h_{27}=1,&h_{26}=2,\\
h_{25}=3,&h_{24}=4,&h_{23}=h_{22}=5,\\
h_{21}=h_{20}=6,&h_{19}=h_{18}=7,&h_{17}=h_{16}=8,\\
h_{15}=h_{14}=h_{13}=9,~~~~&h_{12}=h_{11}=h_{10}=10,
{}~~~~&h_{9}=h_{8}=h_{7}=11,\\
h_{6}=h_{5}=h_{4}=12,~~~~&h_{3}=h_{2}=h_{1}=13,~~~~&h_{\bar{a}}=27-h_{a},
\end{array} \eqno (2.3) $$

In Fig.2, some neighboured blocks are connected by a line. As well
known $^{10}$, in the quantum enveloping algebra $U_{q}E_{7}$ there
are generators $e_{j}$, $f_{j}$ and $k_{j}$, $1\leq j \leq 7$,
satisfying standard quantum algebraic relations $^{10}$. The line
connecting two blocks in Fig.2, labeled a number $j$, denotes that
two states in the blocks can be related by a raising (or lowering)
operator $e_{j}$ (or $f_{j}$). For example, the state $20$ acted by
$e_{2}$ becomes the state $22$. Now, the lines in Fig.2 denotes the
condition that the quantum representation matrix elements of $e_{j}$
and $f_{j}$ related with those two states in the minimal representation
is non-vanishing, namely,
$$D_{q}(e_{j})_{a~b}~=~D_{q}(f_{j})_{b~a}~=~1 \eqno (2.4)$$

\noindent
if $a>b$ and two blocks filled by ${\bf m}_{a}$ and ${\bf m}_{b}$ are
connected by a line labeled by $j$. Otherwise the matrix elements are
vanishing. The quantum representation matrices of $k_{j}$ are diagonal,
and the diagonal elements depend on the weights:
$$D_{q}(k_{j})_{a~a}~=~q^{n_{j}/2},~~~~{\rm if}~~{\bf m}_{a}~=~
\displaystyle \sum_{j=1}^{7}~n_{j}~{\bf \lambda}_{j} \eqno (2.5) $$

The direct product of two minimal representations is a reducible
representation with the Clebsch-Gordan series as follows:
$${\bf \lambda}_{6}~\otimes~{\bf \lambda}_{6}~=~(2{\bf \lambda}_{6})~
\oplus~{\bf \lambda}_{5}~\oplus~{\bf \lambda}_{1}~\oplus~{\bf 0}
\eqno (2.6) $$

\noindent
The dimensions of the representations $(2{\bf \lambda}_{6})$,
${\bf \lambda}_{5}$, ${\bf \lambda}_{1}$, and ${\bf 0}$ are 1463,
1539, 133 and 1, respectively. Their Casimir are 30, 28, 18, and 1,
respectively.

The solution of the simple Yang-Baxter equation related to the
minimal representation of $U_{q}E_{7}$ can be calculated by a
standard method $^{11,10}$:
$$\begin{array}{l}
\breve{R}_{q}~=~{\cal P}_{2{\bf \lambda}_{6}}~-~q^{2}~
{\cal P}_{{\bf \lambda}_{5}}~+~q^{12}~{\cal P}_{{\bf \lambda}_{1}}
{}~-~q^{30}~{\cal P}_{{\bf 0}}\\
\left({\cal P}_{{\bf N}}\right)^{ab}_{~cd}~=~
\left(C_{q}\right)_{{\bf m}_{a}{\bf m}_{b}{\bf N}({\bf m}_{a}+{\bf m}_{b})}
{}~\left(C_{q}\right)_{{\bf m}_{c}{\bf m}_{d}{\bf N}({\bf m}_{a}+{\bf m}_{b})}
\end{array} \eqno (2.7) $$

\noindent
where ${\cal P}_{{\bf N}}$ is a projection operator, and $\left(C_{q}
\right)_{{\bf N}}$ is the $q$-CG matrix reduces
the product representation ${\bf \lambda}_{6} \otimes {\bf \lambda}_{6}$
into the irreducible one ${\bf N}$:
$$| ~{\bf N}, {\bf m}~\rangle~=~\displaystyle \sum_{{\bf m}_{a}}~
|~{\bf \lambda}_{6},{\bf m}_{a}~\rangle
|~{\bf \lambda}_{6},({\bf m}-{\bf m}_{a})~\rangle
{}~\left(C_{q}\right)_{{\bf m}_{a}({\bf m}-{\bf m}_{b}){\bf N}{\bf m}}
\eqno (2.8) $$

\noindent
For convenience, we usually denote by their indices $a$ the states on the
right hand side of (2.8), for example see (3.1).

Our main problems are to calculate the $q$-CG coefficients by (2.8),
and then, to calculate the $\breve{R}_{q}$ matrix by (2.7).

\vspace{10mm}
\noindent
{\bf 3. QUANTUM CLEBSCH-GORDAN COEFFICIENTS}

\vspace{2mm}
The state with the highest weight in the product space is single:
$$| ~2{\bf \lambda}_{6}, 2{\bf \lambda}_{6}~\rangle~=~
|~28~\rangle |~28~\rangle \eqno (3.1) $$

Acting the lowering operator $f_{j}$ on (3.1), we are able to calculate
the expansions for other states. Since the Weyl equivalent states
have the same expansions, we only need to list the expansions
for the states with the dominant weights.

The second dominant weight is ${\bf \lambda}_{5}$. Only two
representations in the CG series (2.6) have this dominant weight:
$$\begin{array}{l}
| ~2{\bf \lambda}_{6}, {\bf \lambda}_{5}~\rangle~=~[2]^{-1/2}~
\left\{q^{1/2}~|~28~\rangle |~27~\rangle
{}~+~q^{-1/2}~|~27~\rangle |~28~\rangle \right\}\\
| ~{\bf \lambda}_{5}, {\bf \lambda}_{5}~\rangle~=~[2]^{-1/2}~
\left\{q^{-1/2}~|~28~\rangle |~27~\rangle
{}~-~q^{1/2}~|~27~\rangle |~28~\rangle \right\}
\end{array} \eqno (3.2) $$

\noindent
where and hereafter we use the following notation as usual:
$$\omega~=~q~-~q^{-1},~~~~~~[n]~=~\omega^{-1}~(q^{n}~-~q^{-n})
 \eqno (3.3)$$

The second expansion in (3.2) is calculated from the orthogonality
of $q$-CG coefficients. Note that the second half terms in the expansions
(3.2) can be obtained from the first half terms by changing the order
of two states and replacing $q$ by $q^{-1}$. The representation
$(2{\bf \lambda}_{6})$ is symmetric so that the terms in the second half
have the same sign as their partner, and the representation
${\bf \lambda}_{5}$ is antisymmetric so that the partners have
opposite signs. In the following expansions we will replace the
second half terms by $(sym.~terms)$ or $(antisym.~terms)$ for simplicity.

Multiplicity appears in the states with the third dominant weight
${\bf \lambda}_{1}$. We distinguish the multiple states by a subscript
$j$ if the state is obtained by the lowering operator $f_{j}$ from a
state with a height lower by one. The expansions of those states are
as follows:
$$\begin{array}{l}
| ~2{\bf \lambda}_{6}, {\bf \lambda}_{1}~\rangle_{3}~=~[2]^{-1}~
\left\{q~ |~25~\rangle |~18~\rangle
{}~+~|~24~\rangle |~20~\rangle ~+~(sym.~terms)\right\}\\
| ~2{\bf \lambda}_{6}, {\bf \lambda}_{1}~\rangle_{2}~=~[2]^{-1}
[3]^{-1/2}~\left\{-q ~|~25~\rangle |~18~\rangle
{}~+~q^{2} ~|~24~\rangle |~20~\rangle \right.\\
\left.~~~~~~~~+~[2] ~|~23~\rangle |~22~\rangle
{}~+~(sym.~terms)\right\}\\
| ~2{\bf \lambda}_{6}, {\bf \lambda}_{1}~\rangle_{4}~=~
([4][3][2])^{-1/2}~\left\{q[3] ~|~26~\rangle |~16~\rangle
{}~+~q^{-2} ~|~25~\rangle |~18~\rangle \right. \\
{}~\left.~~~~~~~-~q^{-1} ~|~24~\rangle |~20~\rangle
{}~+~ |~23~\rangle |~22~\rangle ~+~(sym.~terms)\right\}\\
| ~2{\bf \lambda}_{6}, {\bf \lambda}_{1}~\rangle_{5}~=~
([5][4][2])^{-1/2}~\left\{q[4] ~|~27~\rangle |~13~\rangle~+~
q^{-3} ~|~26~\rangle |~16~\rangle \right.\\
\left.~~~~~~~~-~q^{-2} ~|~25~\rangle |~18~\rangle
{}~+~q^{-1} ~|~24~\rangle |~20~\rangle
{}~-~ |~23~\rangle |~22~\rangle ~+~(sym.~terms)\right\}\\
| ~2{\bf \lambda}_{6}, {\bf \lambda}_{1}~\rangle_{6}~=~
([6][5][2])^{-1/2}~\left\{q[5] ~|~28~\rangle |~10~\rangle~+~
+q^{-4} ~|~27~\rangle |~13~\rangle \right.\\
{}~~~~~~~~-~q^{-3} ~|~26~\rangle |~16~\rangle
{}~+~q^{-2} ~|~25~\rangle |~18~\rangle
{}~-~q^{-1} ~|~24~\rangle |~20~\rangle\\
\left.~~~~~~~~+~ |~23~\rangle |~22~\rangle ~+~(sym.~terms)\right\}
\end{array} \eqno (3.4)$$
$$\begin{array}{l}
| ~{\bf \lambda}_{5}, {\bf \lambda_{1}}~\rangle_{3}~=~[2]^{-1}~
\left\{ |~25~\rangle |~18~\rangle
{}~+~q^{-1}~|~24~\rangle |~20~\rangle ~+~(antisym.~terms)\right\}\\
| ~{\bf \lambda}_{5}, {\bf \lambda_{1}}~\rangle_{2}~=~[2]^{-1}
[3]^{-1/2}~\left\{-~ |~25~\rangle |~18~\rangle
{}~+~q~ |~24~\rangle |~20~\rangle \right.\\
{}~~~~\left.~~~~+~q^{-1}[2]~ |~23~\rangle |~22~\rangle
{}~+~(antisym.~terms)\right\}\\
| ~{\bf \lambda}_{5}, {\bf \lambda_{1}}~\rangle_{7}~=~\left({[2]
\over [4][3]}\right)^{1/2}~\left\{-~ |~25~\rangle |~18~\rangle
{}~+~q ~|~24~\rangle |~20~\rangle \right.\\
\left.~~~~~~~~-~q^{2} ~|~23~\rangle |~22~\rangle
{}~+~(antisym.~terms)\right\}\\
| ~{\bf \lambda}_{5}, {\bf \lambda_{1}}~\rangle_{4}~=~
\left({[2] \over [6][4]}\right)^{1/2}~
\left\{([4]/[2]) ~|~26~\rangle |~16~\rangle
{}~-~q^{-2} \omega ~|~25~\rangle |~18~\rangle \right. \\
{}~\left.~~~~~~~+~q^{-1}\omega ~|~24~\rangle |~20~\rangle
{}~-~\omega ~|~23~\rangle |~22~\rangle ~+~(antisym.~terms)\right\}\\
| ~{\bf \lambda}_{5}, {\bf \lambda_{1}}~\rangle_{5}~=~
\left({[4][3] \over [8][6][2]}\right)^{1/2}~\left\{([6]/[3])
{}~|~27~\rangle |~13~\rangle
{}~-~q^{-3}\omega ~|~26~\rangle |~16~\rangle \right.\\
{}~~~~~~~~+~q^{-2}\omega ~|~25~\rangle |~18~\rangle
{}~-~q^{-1}\omega ~|~24~\rangle |~20~\rangle\\
\left.~~~~~~~~+~\omega ~|~23~\rangle |~22~\rangle
{}~+~(antisym.~terms)\right\}\\
| ~{\bf \lambda}_{5}, {\bf \lambda_{1}}~\rangle_{6}~=~
\left({[5][4]\over [10][8][2]}\right)^{-1/2}~
\left\{([8]/[4]) ~|~28~\rangle |~10~\rangle~
-~q^{-4}\omega ~|~27~\rangle |~13~\rangle \right.\\
{}~~~~~~~~+~q^{-3}\omega ~|~26~\rangle |~16~\rangle
{}~-~q^{-2}\omega ~|~25~\rangle |~18~\rangle
{}~+~q^{-1}\omega ~|~24~\rangle |~20~\rangle\\
\left.~~~~~~~~-~\omega ~|~23~\rangle |~22~\rangle ~+~(antisym.~terms)\right\}
\end{array} \eqno (3.5) $$
$$\begin{array}{l}
| ~{\bf \lambda}_{1}, {\bf \lambda_{1}}~\rangle_{7}~=~\left({[5]
\over [10][6]}\right)^{1/2}~
\left\{q^{-5} ~|~28~\rangle |~10~\rangle~
-~q^{-4} ~|~27~\rangle |~13~\rangle \right.\\
{}~~~~~~~~+~q^{-3} ~|~26~\rangle |~16~\rangle
{}~-~q^{-2} ~|~25~\rangle |~18~\rangle
{}~+~q^{-1} ~|~24~\rangle |~20~\rangle\\
\left.~~~~~~~~-~ ~|~23~\rangle |~22~\rangle ~+~(sym. terms)\right\}
\end{array} \eqno (3.6) $$

The fourth dominant weight is ${\bf 0}$. There are 56 states with
this weight. Here we only list a expansion for a state of the
one-dimensional representation ${\bf 0}$:
$$\begin{array}{l}
| ~{\bf 0}, {\bf 0}~\rangle_{7}=\left({[9][5]
\over [18][14][10]}\right)^{1/2}
\left\{q^{-27/2} ~|28\rangle |\bar{28}\rangle
- q^{-25/2} ~|27\rangle |\bar{27}\rangle
+q^{-23/2} ~|26\rangle |\bar{26}\rangle\right. \\
{}~~~~~-q^{-21/2} ~|25\rangle |\bar{25}\rangle
+q^{-19/2} ~|24\rangle |\bar{24}\rangle
-q^{-17/2} ~|23\rangle |\bar{23}\rangle
-q^{-17/2} ~|22\rangle |\bar{22}\rangle\\
{}~~~~~+q^{-15/2} ~|21\rangle |\bar{21}\rangle
+q^{-15/2} ~|20\rangle |\bar{20}\rangle
- q^{-13/2} ~|19\rangle |\bar{19}\rangle
- q^{-13/2} ~|18\rangle |\bar{18}\rangle\\
{}~~~~~+ q^{-11/2} ~|17\rangle |\bar{17}\rangle
+ q^{-11/2} ~|16\rangle |\bar{16}\rangle
- q^{-9/2} ~|15\rangle |\bar{15}\rangle
- q^{-9/2} ~|14\rangle |\bar{14}\rangle\\
{}~~~~~- q^{-9/2} ~|13\rangle |\bar{13}\rangle
+ q^{-7/2} ~|12\rangle |\bar{12}\rangle
+ q^{-7/2} ~|11\rangle |\bar{11}\rangle
+ q^{-7/2} ~|10\rangle |\bar{10}\rangle\\
{}~~~~~- q^{-5/2} ~|9\rangle |\bar{9}\rangle
- q^{-5/2} ~|8\rangle |\bar{8}\rangle
- q^{-5/2} ~|7\rangle |\bar{7}\rangle
+ q^{-3/2} ~|6\rangle |\bar{6}\rangle\\
{}~~~~~+ q^{-3/2} ~|5\rangle |\bar{5}\rangle
+ q^{-3/2} ~|4\rangle |\bar{4}\rangle
- q^{-1/2} ~|3\rangle |\bar{3}\rangle
- q^{-1/2} ~|2\rangle |\bar{2}\rangle\\
{}~~~~~\left.- q^{-1/2} ~|1\rangle |\bar{1}\rangle
{}~+~(antisym.~terms)\right\}
\end{array} \eqno (3.7) $$

\vspace{10mm}
\noindent
{\bf 4. SOLUTION OF THE YANG-BAXTER EQUATION}

\vspace{2mm}
Now, we are able to calculate the solution $\breve{R}_{q}$ of the
simple Yang-Baxter equation from (2.7). Firstly, from (2.7)
$\breve{R}_{q}$ is a symmetric matrix:
$$\left(\breve{R}_{q}\right)^{ac}_{~bd}~=~
\left(\breve{R}_{q}\right)^{bd}_{~ac} \eqno (4.1) $$

\noindent
Secondly, from the general properties $^{2}$ of $\breve{R}_{q}$, we know:
$$\left(\breve{R}_{q}\right)^{ac}_{~bd}~=~0,~~~~{\rm if}~~c<b
 \eqno (4.2) $$

\noindent
At last, the $q$-CG coefficients given in Sec.3 are invariant
in the Weyl reflection, so do $\breve{R}_{q}$. We only need to
calculate the $\breve{R}_{q}$ matrix elements for the dominant
weights. According to the Weyl orbit sizes of the dominant weights,
we know that the $\breve{R}_{q}$ matrix is a block matrix with
56 1-dimensional submatrices, 756 2-dimensional submatrices,
126 12-dimensional  submatrices, and one 56-dimensional submatrix.

i) $1\times 1$ submatrix for the dominant weight $(2{\bf \lambda}_{6})$

{}From (3.1) we have:
$$\left(\breve{R}_{q}\right)^{28~28}_{~28~28}~=~1
\eqno (4.3) $$

\noindent
{}From the Weyl reflection, (4.3) holds if one replaces 28 by $a$, where
$a$ runs over 28 to $\bar{28}$.

ii) $2\times 2$ submatrix for the dominant weight ${\bf \lambda}_{5}$

{}From (3.2) we have:
$$\left(\breve{R}_{q}\right)^{28~27}_{~27~28}~=~1,~~~~
\left(\breve{R}_{q}\right)^{27~28}_{~27~28}~=~-q~\omega~=~1~-~q^{2}
\eqno (4.4) $$

iii) $12\times 12$ submatrix for the dominant weight ${\bf \lambda}_{1}$

As shown in (3.4) to (3.6), in the direct product space there are
12 states with the dominant weight ${\bf \lambda}_{1}$. They are:
$$\begin{array}{rl}
(a,a')&=~(28,10),~(27,12),~(26,16),~(25,18),~(24,20),~(23,22),\\
&~~~(22,23),~(20,24),~(18,25),~(16,26),~(12,27),~(10,28)
\end{array} \eqno (4.5) $$

\noindent
The rows (columns) of the submatrix with this dominant weight are
also denoted by those pair of numbers $(a,a')$. We would like to
emphasize that the number $a'$ is unique determined by the number
$a$. Now, the calculation results for the submatrix of $\breve{R}_{q}$
are as follows:
$$\left(\breve{R}_{q}\right)^{aa'}_{~bb'}~=~\left\{\begin{array}{ll}
q^{2},&{\rm if}~~h_{a}+h_{b}=10~~{\rm and}~~a\neq b\\
-(-q)^{h_{a}+h_{b}-9}\omega,~~~~&{\rm if}~~h_{a}+h_{b}>10
{}~~{\rm and}~~a\neq b\\
q^{h_{a}-4} \omega^{2} [h_{a}-5],~~~~&{\rm if}~~a=b~~{\rm and}~~
h_{a}\geq 6\\
0, &{\rm the~ rest ~cases} \end{array} \right. \eqno (4.6) $$

\noindent
where the heights $h_{a}$ of the states $a$ are given in (2.3).

iv) $56\times 56$ submatrix for the dominant weight ${\bf 0}$

The rows (columns) of this submatrix are denoted by $(a,\bar{a})$.
The calculation results for the $56\times 56$ submatrix of
$\breve{R}_{q}$ are as follows. There are five cases for the submatrix
elements.

a) If ${\bf m}_{a}+{\bf m}_{b}$ is not a non-positive
combination of the simple roots ${\bf r}_{j}$ of $E_{7}$, i.e.,
there is at least one positive coefficient in the combination,
we have:
$$\left(\breve{R}_{q}\right)^{a\bar{a}}_{~b\bar{b}}~=~0 \eqno (4.7a) $$

b) If $a=-b$, we have:
$$\left(\breve{R}_{q}\right)^{a\bar{a}}_{~\bar{a}a}~=~q^{3} \eqno (4.7b) $$

c) If $a=b$ and ${\bf m}_{a}+{\bf m}_{b}$ is a non-positive
combination of the simple roots ${\bf r}_{j}$ of $E_{7}$, we have:
$$\begin{array}{lll}
\left(\breve{R}_{q}\right)^{\bar{5}5}_{~\bar{5}5}~=~-q^{3}\omega^{3},
&\left(\breve{R}_{q}\right)^{\bar{8}8}_{~\bar{8}8}~=~-q^{4}[2]\omega^{3},
&\left(\breve{R}_{q}\right)^{\bar{9}9}_{~\bar{9}9}~=~-q^{4}[2]\omega^{3},\\
\left(\breve{R}_{q}\right)^{\bar{11}11}_{~\bar{11}11}~=~-q^{5}[3]\omega^{3},
&\left(\breve{R}_{q}\right)^{\bar{12}12}_{~\bar{12}12}~=~-q^{5}[2]^{2}
\omega^{3},
&\left(\breve{R}_{q}\right)^{\bar{13}13}_{~\bar{13}13}~=~-q^{6}[4]
\omega^{3},\\
\left(\breve{R}_{q}\right)^{\bar{14}14}_{~\bar{14}14}~=~-q^{6}[3][2]
\omega^{3},
&\left(\breve{R}_{q}\right)^{\bar{15}15}_{~\bar{15}15}~=~-q^{6}[3][2]
\omega^{3},
&\left(\breve{R}_{q}\right)^{\bar{16}16}_{~\bar{16}16}~=~-q^{7}[4][2]
\omega^{3},\\
\left(\breve{R}_{q}\right)^{\bar{17}17}_{~\bar{17}17}~=~-q^{7}[3]^{2}
\omega^{3},
&\left(\breve{R}_{q}\right)^{\bar{18}18}_{~\bar{18}18}~=~-q^{8}[4][3]
\omega^{3},
&\left(\breve{R}_{q}\right)^{\bar{19}19}_{~\bar{19}19}~=~-q^{8}[4][3]
\omega^{3},\\
\left(\breve{R}_{q}\right)^{\bar{20}20}_{~\bar{20}20}~=~-q^{9}[4]^{2}
\omega^{3},
&\left(\breve{R}_{q}\right)^{\bar{21}21}_{~\bar{21}21}~=~-q^{9}[5][3]
\omega^{3},
&\left(\breve{R}_{q}\right)^{\bar{22}22}_{~\bar{22}22}~=~-q^{10}
[5][4]\omega^{3},\\
\left(\breve{R}_{q}\right)^{\bar{23}23}_{~\bar{23}23}~=~-q^{10}
[5][4]\omega^{3},
&\left(\breve{R}_{q}\right)^{\bar{24}24}_{~\bar{24}24}~=~-q^{11}[5]^{2}
\omega^{3},
&\left(\breve{R}_{q}\right)^{\bar{25}25}_{~\bar{25}25}~=~-q^{12}[6][5]
\omega^{3},\\
\left(\breve{R}_{q}\right)^{\bar{26}26}_{~\bar{26}26}~=~-q^{13}[7][5]
\omega^{3},
&\left(\breve{R}_{q}\right)^{\bar{27}27}_{~\bar{27}27}~=~-q^{14}[8][5]
\omega^{3},
&\left(\breve{R}_{q}\right)^{\bar{28}28}_{~\bar{28}28}~=~-q^{15}[9][5]
\omega^{3},
\end{array} \eqno (4.7c) $$

d) If ${\bf m}_{a}+{\bf m}_{b}$ is a negative root of $E_{7}$,
we have:
$$\left(\breve{R}_{q}\right)^{a\bar{a}}_{~b\bar{b}}~=~(-q)^{h_{a}+h_{b}-25}
\omega \eqno (4.7d) $$

e) If $a\neq \pm b$, and ${\bf m}_{a}+{\bf m}_{b}$ is a
non-positive combination of the simple roots ${\bf r}_{j}$, but not a
negative root of $E_{7}$, we have:
$$\left(\breve{R}_{q}\right)^{a\bar{a}}_{~b\bar{b}}~=~
(-q)^{h_{a}+h_{b}-25}q^{-\alpha} [\alpha] \omega^{2} \eqno (4.7e) $$

\noindent
where the following pairs $(a,b)$ correspond to $\alpha=1$:
$$\begin{array}{llllllll}
(\bar{5},\bar{1}),&(\bar{5},\bar{2}),&(\bar{5},\bar{3}),&(\bar{8},1),
&(\bar{8},\bar{2}),&(\bar{8},\bar{3}),&(\bar{8},\bar{5}),&(\bar{9},3),\\
(\bar{9},\bar{1}),&(\bar{9},\bar{2}),&(\bar{9},\bar{5}),&(\bar{11},4),
&(\bar{11},1),&(\bar{11},\bar{2}),&(\bar{11},\bar{3}),&(\bar{11},\bar{5}),\\
(\bar{11},\bar{8}),&(\bar{12},5),&(\bar{12},3),&(\bar{12},1),
&(\bar{12},\bar{2}),&(\bar{13},7),&(\bar{13},4),&(\bar{13},1),\\
(\bar{13},\bar{2}),&(\bar{13},\bar{3}),&(\bar{13},\bar{5}),&(\bar{13},\bar{8}),
&(\bar{13},\bar{11}),&(\bar{14},8),&(\bar{14},5),&(\bar{14},4),\\
(\bar{14},3),&(\bar{14},1),&(\bar{14},\bar{2}),&(\bar{16},11),
&(\bar{16},8),&(\bar{16},7),&(\bar{16},5),&(\bar{16},4),\\
(\bar{16},3),&(\bar{16},1),&(\bar{16},\bar{2}),&(\bar{17},12),
&(\bar{17},8),&(\bar{17},4),&(\bar{18},14),&(\bar{18},12),\\
(\bar{18},11),&(\bar{18},8),&(\bar{18},7),&(\bar{18},4),
&(\bar{20},17),&(\bar{20},14),&(\bar{20},11),&(\bar{20},7),\\
(\bar{23},19),&(\bar{23},17),&(\bar{23},14),&(\bar{23},11),
&(\bar{23},7),&(\bar{24},20),&(\bar{24},19),&
\end{array} $$

\noindent
the following pairs $(a,b)$ correspond to $\alpha=2$:
$$\begin{array}{llllllll}
(\bar{8},\bar{4}),&(\bar{9},\bar{6}),&(\bar{12},\bar{4}),&(\bar{12},\bar{6}),
&(\bar{12},\bar{8}),&(\bar{12},\bar{9}),&(\bar{14},\bar{6}),
&(\bar{14},\bar{9}),\\
(\bar{14},\bar{12}),&(\bar{15},2),&(\bar{15},\bar{1}),&(\bar{15},\bar{3}),
&(\bar{15},\bar{4}),&(\bar{15},\bar{5}),&(\bar{15},\bar{6}),
&(\bar{15},\bar{8}),\\
(\bar{15}),\bar{9}),&(\bar{15},\bar{12}),&(\bar{16},\bar{6}),
&(\bar{16},\bar{9}),&(\bar{16},\bar{12}),&(\bar{16},\bar{14}),
&(\bar{17},2),&(\bar{17},\bar{1}),\\
(\bar{17},\bar{3}),&(\bar{17},\bar{5}),&(\bar{17},\bar{6}),
&(\bar{17},\bar{9}),&(\bar{18},2),&(\bar{18},\bar{1}),&(\bar{18},\bar{3}),
&(\bar{18},\bar{5}),\\
(\bar{18},\bar{6}),&(\bar{18},\bar{9}),&(\bar{19},9),&(\bar{19},5),
&(\bar{19},2),&(\bar{19},1),&(\bar{19},\bar{3}),&(\bar{19},\bar{6}),\\
(\bar{20},9),&(\bar{20},5),&(\bar{20},2),&(\bar{20},1),
&(\bar{20},\bar{3}),&(\bar{20},\bar{6}),&(\bar{22},15),&(\bar{22},12),\\
(\bar{22},9),&(\bar{22},8),&(\bar{22},5),&(\bar{22},4),
&(\bar{22},2),&(\bar{22},1),&(\bar{22},\bar{3}),&(\bar{22},\bar{6}),\\
(\bar{24},15),&(\bar{24},12),&(\bar{24},8),&(\bar{24},4),
&(\bar{25},18),&(\bar{25},17),&(\bar{25},15)&
\end{array} $$

\noindent
the following pairs $(a,b)$ correspond to $\alpha=3$:
$$\begin{array}{llllllll}
(\bar{11},\bar{7}),&(\bar{14},\bar{7}),&(\bar{14},\bar{11}),
&(\bar{17},\bar{7}),&(\bar{17},\bar{11}),&(\bar{17},\bar{14}),
&(\bar{17},\bar{15}),&(\bar{18},\bar{15}),\\
(\bar{18},\bar{17}),&(\bar{19},\bar{4}),&(\bar{19},\bar{7}),
&(\bar{19},\bar{8}),&(\bar{19},\bar{11}),&(\bar{19},\bar{12}),
&(\bar{19},\bar{14}),&(\bar{19},\bar{15}),\\
(\bar{19}),\bar{17}),&(\bar{20},\bar{4}),&(\bar{20},\bar{8}),
&(\bar{20},\bar{12}),&(\bar{20},\bar{15}),&(\bar{21},6),&(\bar{21},3),
&(\bar{21},\bar{1}),\\
(\bar{21},\bar{2}),&(\bar{21},\bar{4}),&(\bar{21},\bar{5}),
&(\bar{21},\bar{7}),&(\bar{21},\bar{8}),&(\bar{21},\bar{9}),
&(\bar{21},\bar{11}),&(\bar{21},\bar{12}),\\
(\bar{21},\bar{14}),&(\bar{21},\bar{15}),&(\bar{21},\bar{17}),
&(\bar{21},\bar{19}),&(\bar{23},6),&(\bar{23},3),&(\bar{23},\bar{1}),
&(\bar{23},\bar{2}),\\
(\bar{23},\bar{4}),&(\bar{23},\bar{5}),&(\bar{23},\bar{8}),
&(\bar{23},\bar{9}),&(\bar{23},\bar{12}),&(\bar{23},\bar{15}),
&(\bar{24},6),&(\bar{24},3),\\
(\bar{24},\bar{1}),&(\bar{24},\bar{2}),&(\bar{24},\bar{5}),
&(\bar{24},\bar{9}),&(\bar{25},9),&(\bar{25},6),&(\bar{25},5),&(\bar{25},3),\\
(\bar{25},1),&(\bar{25},\bar{2}),&(\bar{26},16),&(\bar{26},14),
&(\bar{26},12),&(\bar{26},9),&(\bar{26},6),&
\end{array} $$

\noindent
the following pairs $(a,b)$ correspond to $\alpha=4$:
$$\begin{array}{llllllll}
(\bar{13},\bar{10}),&(\bar{16},\bar{10}),&(\bar{16},\bar{13}),
&(\bar{18},\bar{10}),&(\bar{18},\bar{13}),&(\bar{18},\bar{16}),
&(\bar{20},\bar{10}),&(\bar{20},\bar{13}),\\
(\bar{20},\bar{16}),&(\bar{20},\bar{18}),&(\bar{20},\bar{19}),
&(\bar{22},\bar{7}),&(\bar{22},\bar{10}),&(\bar{22},\bar{11}),
&(\bar{22},\bar{13}),&(\bar{22},\bar{14}),\\
(\bar{22},\bar{16}),&(\bar{22},\bar{17}),&(\bar{22},\bar{18}),
&(\bar{22},\bar{19}),&(\bar{22},\bar{20}),
&(\bar{23},\bar{10}),&(\bar{23},\bar{13}),&(\bar{23},\bar{16}),\\
(\bar{23},~\bar{18}),&(\bar{23},\bar{20}),
&(\bar{24},\bar{7}),&(\bar{24},\bar{10}),&(\bar{24},\bar{11}),
&(\bar{24},\bar{13}),&(\bar{24},\bar{14}),&(\bar{24},\bar{16}),\\
(\bar{24},\bar{17}),&(\bar{24},\bar{18}),
&(\bar{25},\bar{4}),&(\bar{25},\bar{7}),&(\bar{25},\bar{8}),
&(\bar{25},\bar{10}),&(\bar{25},\bar{11}),&(\bar{25},\bar{12}),\\
(\bar{25},\bar{13}),&(\bar{25},\bar{14}),&(\bar{25},\bar{16}),
&(\bar{26},2),&(\bar{26},\bar{1}),&(\bar{26},\bar{3}),
&(\bar{26},\bar{4}),&(\bar{26},\bar{5}),\\
(\bar{26},\bar{7}),&(\bar{26},\bar{8}),&(\bar{26},\bar{10}),
&(\bar{26},\bar{11}),&(\bar{26},\bar{13}),&(\bar{27},13),
&(\bar{27},11),&(\bar{27},8),\\
(\bar{27},5),&(\bar{27},3),&(\bar{27},2),&(\bar{27},\bar{1}),
&(\bar{27},\bar{4}),&(\bar{27},\bar{7}),&(\bar{27},\bar{10}),&
\end{array} $$

\noindent
and at last, the following pairs $(a,b)$ correspond to $\alpha=5$:
$$\begin{array}{llllllll}
(\bar{23},\bar{21}),&(\bar{24},\bar{21}),&(\bar{24},\bar{22}),
&(\bar{24},\bar{23}),&(\bar{25},\bar{19}),&(\bar{25},\bar{20}),
&(\bar{25},\bar{21}),&(\bar{25},\bar{22}),\\
(\bar{25},\bar{23}),&(\bar{25},\bar{24}),&(\bar{26},\bar{15}),
&(\bar{26},\bar{17}),&(\bar{26},\bar{18}),&(\bar{26},\bar{19}),
&(\bar{26},\bar{20}),&(\bar{26},\bar{21}),\\
(\bar{26},\bar{22}),&(\bar{26},\bar{23}),&(\bar{26},\bar{24}),
&(\bar{26},\bar{25}),&(\bar{27},\bar{6}),&(\bar{27},\bar{9}),
&(\bar{27},\bar{12}),&(\bar{27},\bar{14}),\\
(\bar{27},\bar{15}),&(\bar{27},\bar{16}),&(\bar{27},\bar{17}),
&(\bar{27},\bar{18}),&(\bar{27},\bar{19}),&(\bar{27},\bar{20}),
&(\bar{27},\bar{21}),&(\bar{27},\bar{22}),\\
(\bar{27},\bar{23}),&(\bar{27},\bar{24}),&(\bar{27},\bar{25}),
&(\bar{27},\bar{26}),&(\bar{28},10),&(\bar{28},7),
&(\bar{28},4),&(\bar{28},1),\\
(\bar{28},\bar{2}),&(\bar{28},\bar{3}),&(\bar{28},\bar{5}),
&(\bar{28},\bar{6}),&(\bar{28},\bar{8}),&(\bar{28},\bar{9}),
&(\bar{28},\bar{11}),&(\bar{28},\bar{12}),\\
(\bar{28},\bar{13}),&(\bar{28},\bar{14}),&(\bar{28},\bar{15}),
&(\bar{28},\bar{16}),&(\bar{28},\bar{17}),&(\bar{28},\bar{18}),
&(\bar{28},\bar{19}),&(\bar{28},\bar{20}),\\
(\bar{28},\bar{21}),&(\bar{28},\bar{22}),&(\bar{28},\bar{23}),
&(\bar{28},\bar{24}),&(\bar{28},\bar{25}),&(\bar{28},\bar{26}),
&(\bar{28},\bar{27}),&
\end{array}  $$

The first three submatrices of $\breve{R}_{q}$ were given in Ref.9,
but the $56\times 56$ submatrix is firstly obtained here.

Because we have obtained all the $q$-CG coefficients for the
direct product of two minimal representation, it is easy to
calculate the spectrum-dependent solution of the Yang-Baxter
equation by the standard method $^{4,10}$. From (7.61) of Ref.10
we have:
$$\begin{array}{rl}
\breve{R}_{q}(x)&=~(1-xq^{2})(1-xq^{10})(1-xq^{18})~{\cal P}_{2{\bf
\lambda}_{6}}
{}~+~(x-q^{2})(1-xq^{10})(1-xq^{18})~{\cal P}_{{\bf \lambda}_{5}}\\
&~~~+~(x-q^{2})(x-q^{10})(x-q^{18})~{\cal P}_{{\bf \lambda}_{1}}
{}~+~(x-q^{2})(x-q^{10})(x-q^{18})~{\cal P}_{{\bf 0}}
\end{array} \eqno (4.8) $$

\noindent
It coincides with the solution given in Ref.12.

By the way, in the theory of quantum groups the $\hat{R}_{q}$
matrix is usually chosen as follows:
$$\hat{R}_{q}~=~q~\breve{R}_{q}^{-1} \eqno (4.9) $$

\noindent
where
$$\left(\breve{R}_{q}^{-1}\right)^{ac}_{~bd}~=~
\left. \left(\breve{R}_{q}\right)^{ca}_{~db_{~_{~}}}
\right|_{q\rightarrow q^{-1}} \eqno (4.10) $$

\vspace{2.0cm}
{\bf Acknowledgments}. This work was supported by the National
Natural Science Foundation of China and Grant No. LWTZ-1298 of
Chinese Academy of Sciences.

\end{document}